\documentclass[twocolumn]{revtex4}
\usepackage{graphicx}
\usepackage[]{epsfig}

\newcommand{\beq}{\begin{equation}}
\newcommand{\eeq}{\end{equation}}
\newcommand{\beqd}{\begin{displaymath}}
\newcommand{\eeqd}{\end{displaymath}}
\newcommand{\beqa}{\begin{eqnarray}}
\newcommand{\eeqa}{\end{eqnarray}}
\newcommand{\non}{\nonumber}

\newcommand{\comment}[1]{}

\begin{document}
\title{Large Deviations in the Free-Energy of Mean-Field Spin-Glasses}


\author{Giorgio Parisi$^{1}$ and Tommaso Rizzo$^{2}$}

\affiliation{$^{1}$Dipartimento di Fisica, Universit\`a di Roma ``La Sapienza'', 
P.le Aldo Moro 2, 00185 Roma,  Italy
\\
$^{2}$ ``E. Fermi'' Center, Via Panisperna 89 A, Compendio Viminale, 00184, Roma, Italy}

\begin{abstract}
    We compute analytically the probability distribution of large deviations in the spin-glass free energy 
     for the Sherrington-Kirkpatrick mean field model, {\it i.e.} we compute the exponentially small
    probability of finding a system with intensive free energy smaller than the most likely one.  This result
    is obtained by computing the average value of the partition function to the power $n$ as a function of $n$.  At
    zero temperature this absolute prediction displays a remarkable quantitative agreement with the numerical
    data.

\end{abstract} 

\maketitle

In the study of disordered systems nearly all predictions concern the most likely behavior, but there is
also considerable interest in developing techniques to compute the probability distribution of
rare events, i.e. the probability  of finding systems that have properties different
from the typical ones. The motivations are various:
\begin{itemize}
    \item Systems with behavior different form the most likely one may have some special interest.
    \item The comparison between analytic predictions in  the large deviations region and numerical or
    experimental data may provide a clear-cut test of the theoretical approach used to compute the most likely
    properties.
    \item The properties of large fluctuations may be related to other more interesting properties of the 
    system.
\end{itemize}

Unfortunately even in the simplest non-trivial case, i.e. the Sherrington-Kirkpatrick (SK) infinite range
model for spin glasses, there is no consensus on the procedure to perform such a computation. Everybody agrees
that as a first step we need to compute the thermodynamic function
\beq
\Phi(n,\beta)=-{1\over \beta n N}\ln \overline{Z_{J}(\beta)^n}\ ,
\eeq
where different systems (or samples) are labeled by $J$, $Z_{J}(\beta)$ is the partition function and the
bar denotes the average over different disordered samples.  Indeed it is well known that the probability of large
deviations is related to the function $\Phi(n,\beta)$. 

The disagreement is in the
computation of  $\Phi(n,\beta)$. At $n=0$ it can be done using the approach of broken replica symmetry (that is
known to give the exact results), where it coincides with the most likely free energy $\Phi(0,\beta)=f_{typ}$ or equivalently with the average equilibrium free energy $f_{eq}=f_{typ}$. 

For $n>0$ Kondor \cite{Kon1} in 1983 presented a first computation of $\Phi(n,\beta)$ in the region near
$T_{c}$ using the most natural ansatz for replica symmetry breaking (RSB) obtaining $\Phi(n,\beta)= f_{typ}+An^5$.
However it was not possible to test directly Kondor prediction because all numerical data concern the
fluctuations of the ground state energy, i.e. the system is at zero temperature. 

Many efforts has been concentrated on the
scaling of the small deviations of the free energy.  Indeed based on Kondor's result it was argued in \cite{CPSV}
that the small deviations of the free energy per spin from its mean scale as $N^{-5/6}$.  This prediction has
been put to test in a series of numerical works \cite{B1,BKM,CMPP,PALA,B2,KKLJH,PAL} and although all 
estimates are smaller than $5/6$ nobody has claimed that this value is definitively ruled out.  However it was
difficult to test the theory in absence of a quantitative prediction (the only prediction being on the
exponent, a quantity that it is rather difficult to measure in a reliable way).

More recently a
different replica symmetry breaking ansatz was proposed by Aspelmeier and Moore
\cite{AM,DDF}, who found $\Phi(n)=f_{typ}$; in their approach the probability of large deviations goes to zero
faster than $\exp(-\Delta\Sigma(f) N)$ and the small deviations has to be computed with an approach that is not related to
large deviations.

In this letter we concentrate on large deviations. We follow Kondor's approach and we
extend his computation to all temperatures, including $T=0$; in this way we obtain an absolute
prediction for the large deviations distribution. We compare our analytic results with the
numerical simulations done at zero temperature and we find a remarkable agreement.We also
find  that the
alternative approach \cite{AM,DDF} cannot be valid for large positive $n$ and there are no compelling reasons for which it 
should be valid at fixed positive $n$ when $N$ goes to infinity. The problem of computing the large deviations for the SK model at all
temperatures is thus solved.

We start our analysis by defining the sample complexity $\Delta \Sigma(f)$ as the logarithm divided by $N$
of the probability density of finding a  sample of size $N$ with free energy per spin $f$ in the
thermodynamic limit \cite{CPSV}, i.e.
\beq
\Delta \Sigma(f) =\lim_{N\to\infty}{\log(P_{N}(f)) \over N }\ .
\eeq
For large $N$ the majority
of the samples has free energy per spin equal to $f_{typ}$, and all other values have exponentially small
probability.  Consistently $\Delta \Sigma(f)$ is less or equal than zero, the equality holding for $f=f_{typ}$,
i.e. $\Delta \Sigma(f_{typ})=0$. For some values of $f$ it is possible that $\Delta \Sigma(f)=-\infty$,
signalling that the probability of large deviations goes to zero faster than exponentially with $N$.

It is evident that $\Phi(n)$ is the Legendre transform of $\Delta \Sigma(f)$:
\beq
-\beta n \Phi(n)=- \beta n f+ \Delta \Sigma(f)\ , \ \ \ \beta n={\partial \Sigma \over \partial f}
\label{prima}
\eeq
Equivalently we have:
\beq
 \Delta \Sigma(f)=\beta n f-\beta n \Phi(n) \ , \ \ \ f={\partial (n \Phi(n)) \over \partial n} .
\eeq
\begin{figure}[htb]
\begin{center}
\epsfig{file=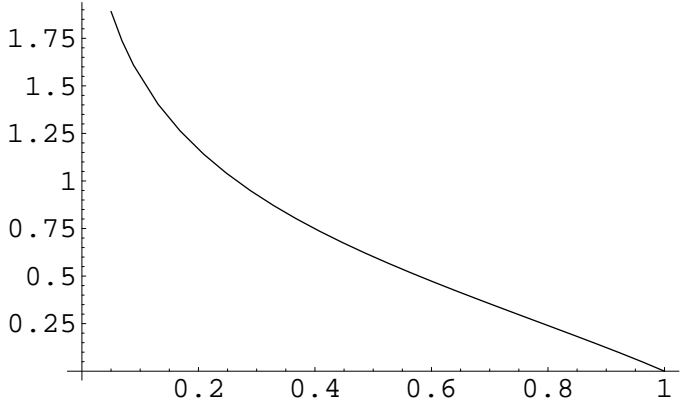,width=9cm}
\caption{The dAT line in the $(T,\beta n)$ plane.  The value of $\beta n$ diverges in the zero-temperature
limit as $\beta n\simeq \sqrt{-2 \ln T}$, as a consequence the function $\Delta \Sigma(\Delta e)$ at zero
temperature is described by the RSB solution at any value of $\Delta e$}
\label{dAT}
\end{center}\end{figure}

Generally speaking for positive $n$ one must distinguish two regions in the $T-n$ plane separated by the so called
de Almeida Thouless (dAT) line, see fig. (\ref{dAT}). In the region above the dAT line, the phase is replica-symmetric, while replica symmetry is broken below.

In the Replica-Symmetric (RS) region the order parameter is the overlap $q$.  The corresponding value of the
potential $\Phi(n,q)$ is given by
\beqa
    F_{n}(q) & = &  -{\beta \over 4}\left(1- 2 q+(1-n) q^2\right)
+
\non
\\
& - & {1 \over \beta n}\ln \int_{-\infty}^{+\infty}{dy \over \sqrt{2 \pi q}}e^{-{y^2\over 2
 q}}(2\cosh \beta y)^n \ .
\non
\eeqa
The overlap $q$ can be computed by solving the  equation $\partial \Phi(n,q)/ \partial q =0$, that is
equivalent to :
\beq
q={ 
\int e^{-{y^2\over 2 q}}(\cosh \beta y)^n \tanh^2 \beta y \, dy
\over
\int e^{-{y^2\over 2 q}}(\cosh \beta y)^n \, dy
}
\label{qRS}
\eeq

In the low temperature phase the RS solution is unstable at small values of $n$.  In the $(n,T)$ plane
the dAT line is specified by the condition \cite{MPV}:
\beq
T^2={ 
\int e^{-{y^2\over 2 q}}(\cosh \beta y)^n (1-\tanh^2 \beta y)^2 \, dy
\over
\int e^{-{y^2\over 2 q}}(\cosh \beta y)^n \, dy
}
\eeq
On the dAT line the value of $n$ is $n_{dAT}(T)= 4 \tau /3$ for small $\tau=1-T$ while $n_{dAT}(T)$ vanishes in the
zero-temperature limit as $n=T\sqrt{-2 \ln T}$.  As a consequences in the rescaled $(T, n \beta)$ plane the dAT line
never touches the $T=0$ line and $\Delta \Sigma(e)$ at $T=0$ is always in the RSB phase, see fig (\ref{dAT}).
 
For small $n$ the RS solution is not only unstable but also inconsistent, indeed near the critical temperature 
(for $T<T_{c}$) we find an unphysical positive value of the complexity difference. 
At any finite temperature $\Phi(n)$ is described by the RS solution at large values of $n$.  Both above and
below the critical temperature, the behavior of $\Phi(n)$ for large values of $n$ is 
$\Phi(n)=-\beta n/4-\ln 2/(\beta n)+O(e^{-2\beta n})$.  This leads to $\Delta \Sigma(f)=-f^2+\ln 2+o(1)$ 
for large negative $f$, note that this is the same behavior of the Random-Energy-Model \cite{REM}.

\begin{figure}[htb]
\begin{center}
\epsfig{file=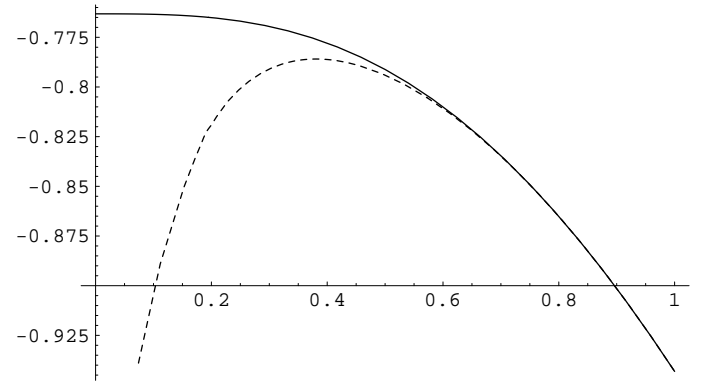,width=9cm}
\caption{Free energy vs.  temperature at equilibrium (solid) \cite{CR} and on the dAT line (dashed), for small
$\tau=T_c-T$ the difference is $f_{eq}-f_{dAT}=2 \tau^5/45+O(\tau^6)$.  The free energy on the dAT line
diverges as $-\sqrt{-(\ln T)/2}$ at low
temperatures.  }
\label{fdat}
\end{center}\end{figure}

Below the dAT line we must break the replica symmetry.  Since the free energy at the dAT line is not equal
to the most likely free energy ($f_{typ}$), see fig. (\ref{fdat}), we must look for a free energy that shows some dependence on $n$
also in this region and the one suggested by Kondor is the most natural one.

In Kondor's approach  for $n<n_{dAT}(T)<1$  one introduces a function $q(x)$ defined
for $n\le x\le 1$ that describes the breaking of replica symmetry in the low temperature phase; A functional $F_{n}[q]$ is obtained such that
$\Phi(n)=\max_{q}F_{n}[q]$.  The function $q(x)$ that maximizes $F_{n}[q]$ can be found by solving the
stationarity equation $\delta F / \delta q(x) = 0$.  This generalizes the standard approach that is proved
to give the correct value of $\Phi(n)$ at $n=0$ \footnote{For larger values of $n$ Talagrand \cite{TALA} was
able to show rigorously that this approach gives a lower bound to the exact results.}.

The form of the free energy functional is the usual one, the only difference being that all functions are defined
in the interval $n\le x \le 1$. One finds that
\begin{displaymath}
    F_{n}[q]  =  -{\beta \over 4}\left(1- 2 q(1)+\int_n^1  q^2(x)\, dx\right)
+
\end{displaymath}
\begin{displaymath}
 - {1 \over \beta n}\ln \int_{-\infty}^{+\infty}{dy \over \sqrt{2 \pi q(n)}}\exp \left(-{(y-h)^2\over 2
 q(n)}\right)\exp(\beta n f(n,y)) \ .
\end{displaymath}
The function $f(x,y)$ is defined in the strip $n\le x \le 1$ and obeys the following equation:
\beq
\dot{f}=-{\dot{q} \over 2}(f''+\beta x (f')^2)\ .
\label{eqf}
\eeq
where dots and primes mean respectively derivatives with respect to $x$ and $y$. The initial condition is on
the right boundary of the strip, at $x=1$, where $
\beta f(1,y)=\log 2 \cosh \beta y $.
There are many ways in which one can compute the maximum of $F_n[q]$. Here we follow
\cite{SD,CR} and introduce  Lagrange multipliers $P(x,y)$ to enforce 
equation (\ref{eqf}). The resulting equations are:
\beq
q(x)=\int_{-\infty}^{\infty}P(x,y) m^2(x,y) dy
\eeq
\beq
m=f'\, ; \ \ \ \dot{m}=-{\dot{q}\over 2}(m''+2 x \beta \,m \,m' )
\eeq
\beq
\dot{P}={\dot{q}\over 2}(P''-2 x \beta \,(m \,P)' )
\eeq
These are the same equations of the standard $n\rightarrow 0$ case, the only difference being in the initial
condition for $P(x,y)$ that reads:
\beq
P(0,y)= c \, \exp \left[ -{(y-h)^2 \over 2 \, q(n)}+ \beta \, n \, f(n,y)  \right]
\eeq
where $c$ is a normalization constant in order to have $\int P(x,y) dy=1$.  Since $ F_{n}[q]$ is extremized with
respect to $q(x)$, the conjugate variable $f$ can be obtained as the partial derivative of $n  F_{n}[q]$ with respect to $n$ evaluated
at the saddle point:
\beq
f=-{\beta \over 4}\left(1- 2 q(1)+\int_n^1  q^2(x) \, dx -n q^2(n) \right)-\langle f(n,y) \rangle \ ,
\eeq  
where square brackets represent average with respect to the measure $ d \mu = \exp (-(y-h)^2 / 2 q(n) + \beta n f(n,y))$.

Kondor \cite{Kon1} found that near the critical
temperature $\Phi(n)=f_{typ}-9 n^5/5120$.  We have solved the RSB equations and computed $q(n,x)$ and $\Phi(n)$ as a series in powers of $n$ and $\tau=1-T$
\cite{CR} up to the 18th order, the series is reported in appendix. It can be proved (and it is confirmed by the
explicit computation) that  the lowest power of $n$ in the expansion of $\Phi(n)$ is $n^5$ and that there is no $n^6$ term.
For negative $n$ the saddle point of the $F_{n}[q]$ is the standard $q(x)$ corresponding to $n=0$, thus
$\Phi(n)=f_{typ}$ for $n<0$ \cite{DFM}.  The corresponding sample complexity as a function of $\Delta f=f-f_{typ}$ reads:
\beqa
\Delta \Sigma(f) &  = & -\infty \  \ \ \ \ \ \ \ \ \ \ \ \ \ \ \ \ \ \ \ \ \ \ \ \ \ \ \ \ \ \   {\rm for}\  \Delta f>0
\nonumber
\\ 
\Delta \Sigma(f) &  = & a_{6/5} |\Delta f|^{6 / 5} +O(|\Delta f|^{8 / 5})  \ \ \ {\rm for} \ \Delta f \leq 0
\nonumber
\eeqa
Where $a_{6/5}=-5\beta  |c_5|^{-{1 / 5}} 6^{-{6 / 5}}$ and $c_5$ is the coefficient of $n^5$ in the expansion of $\Phi(n)$.

We have verified  by an expansion in
powers of $n$ at {\it finite} temperature that {\it the $O(n^5)$ scaling of $\Phi(n)$  holds true at all temperatures} as follows from an  analytic argument that for reasons of space will be reported elsewhere \cite{PRprep}.

It is interesting to note that from the third order on, all derivatives of $\Phi(n)$ (with respect to $n$,$T$
and both) are discontinuous on the dAT line {\it i.e.} the transition is third order.  This is the same
behavior of the free energy on the dAT line in the $(h,T)$ plane \cite{CRT}.

When $\beta \rightarrow \infty$ the complexity $\Delta \Sigma(f)$ goes to a well-defined limit. Therefore from
eq.  (\ref{prima}) $\Phi(n)$ is actually a function of $\beta n$ and the coefficient $c_a$ of
$n^a$ in the power series of $\Phi(n)$ diverges as $\beta^a$ in the zero temperature limit.

The series in powers of $\tau$ of $c_5$ (the $n^5$ coefficient in $\Phi(n)$) can be used to obtain its behavior
in the whole low temperature phase provided one uses the information that $c_5 \sim \beta^5$ in the
zero-temperature limit.  Indeed the series can be resummed using  Pad\'e
approximants with estimated errors not greater that 1\% in the whole temperature range.

\begin{figure}[htb]
\begin{center}
\epsfig{file=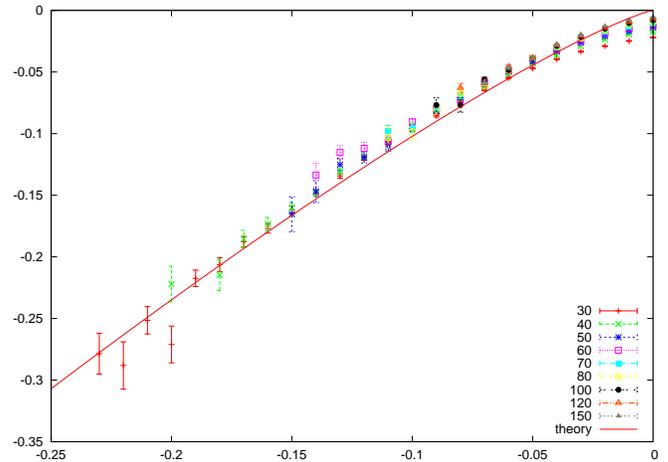,width=9cm}
\caption{Comparison between the numerical and analytical sample complexity at zero temperature, see text.  The
data are those of Ref.  \cite{CMPP}.  }
\label{numeric}
\end{center}\end{figure}

From the Pad\'e approximants of $c_5 \beta^{-5}$ and $c_7 \beta^{-7}$ we estimate $c_5 \simeq -0.0060(1) \,
\beta^5 $ near $T=0$ and $c_7 \simeq -0.0150(5) \beta^7$ in the SK model.
The zero temperature complexity then reads:
\beq
\Delta \Sigma=-1.62(1) \, |\Delta e|^{6/5}+3.1(1)\, |\Delta e|^{8/5}+O(\Delta e^{8/5})\ ,
\label{DSx}
\eeq

Unfortunately the second term yields a big correction to the first one, indeed: i) the exponents of the
series grow slowly (as $(6+i)/5$, $i=0,2,3,\dots$ because there is no $n^6$ term in $\Phi(n)$) and ii) the
coefficients of the series grow quickly with order, actually we expect the series to be asymptotic as is
usually the case in this context \cite{CR}.

In order to bypass this problem and have a good control on $\Delta \Sigma(\Delta e)$ we have adopted a method introduced in
\cite{CR} to obtain $q(x,\tau)$ from its series in powers of $x$ and $\tau$.  We have transformed the series
of $\Delta \Sigma(\Delta f)$ in powers of $\Delta f$ and $\tau$ in a power series of just $\tau$ by setting
$\Delta f=({2 \over 45}s^5+{1\over 4}\tau^7)c$ with $c$ a parameter in the range $[0,1]$.  The corresponding
series in powers of $\tau$ were resummed for any given $c$ through Pad\'e approximants obtaining the curve
$\Delta \Sigma(\Delta e)$ in parametric form.
By resumming the series of $\Phi(n,\tau)$ as a function of $\tau$ we have been able to obtain the sample complexity in the whole
low-temperature phase using the technique of Pad\'e approximants: 18 orders  of the Taylor expansion give us  
a very good control on the function. 

\begin{figure}[htb]
\begin{center}
\epsfig{file=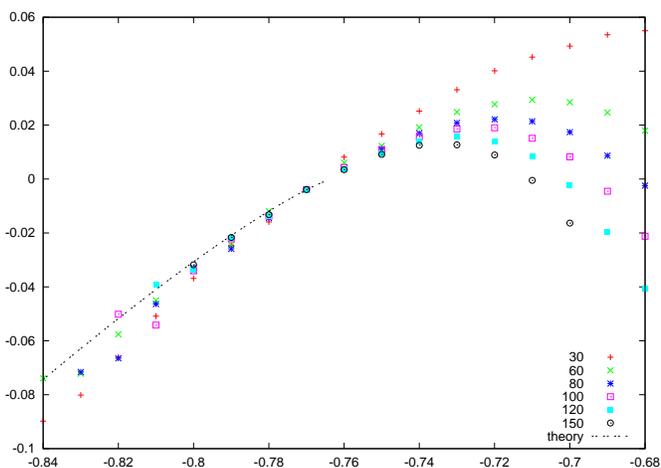,width=9cm}
\caption{Plot of the numerical complexity (from \cite{CMPP}) as
a function of the  energy density for different sample sizes
at zero temperature.  The data have been shifted vertically by an amount
$\Delta_N$ so that the complexity vanishes at the typical energy
$E_{typ}=-.7633$.  }
\label{absolute}
\end{center}\end{figure}

Using this technique  we find that, for not too large $\Delta e$, the zero temperature result
shown in fig.  (\ref{numeric}) differs by less than $1\%$  from the first term of eq.(\ref{DSx}) in this range of energy differences.
We compared the sample complexity with the numerical data at zero temperature of Ref.  \cite{CMPP} as a function
$\Delta e$ and find a very good agreement.  For each $N$ we have plotted $\Delta \Sigma_N=\ln(P(\Delta
e_N)/N^{5/6})/N$ with $\Delta e_N=e-e_N$ (the average energy at size $N$): we put
the $N^{5/6}$ factor in the definition so that that
 $\Delta \Sigma$  goes to a constant for $\Delta e_N=0$.  
Unfortunately the dependence of $e_{N}$ on the size is not negligible and the reader may wonder what happens if
we plot the data at fixed energy.  This has been done in fig.(\ref{absolute}), where data (from \cite{CMPP})
have been shifted vertically by an amount $\Delta_N$ so that the complexity vanishes at the typical energy
$E_{typ}=-.7633$, as it should do in the infinite volume limit.  This correction ($\Delta_N$) goes to zero at
large $N$ but it is important for having a good scaling for finite $N$.  It is interesting to note that for
$E<E_{typ}$ the numerical data approach the theoretical prediction from below, thus strongly suggesting that $\Delta
\Sigma(\Delta e)$ is finite 
at variance with the alternative scenario where
$\Delta\Sigma$ goes to $-\infty$ when $N$ goes to infinity.

Using the standard hierarchical ansatz we have computed the large deviations function at all temperatures.  In
this way we have been able to confirm that the sample complexity $\Delta \Sigma(\Delta f)$ is proportional to
$|\Delta f|^{6/5}$ for small negative $\Delta f$, this result strongly suggests that the sample-to-sample
fluctuations are proportional to $N^{-5/6}$.  We have verified that the numerical data of \cite{CMPP} are in
remarkably good agreement with our absolute prediction. We believe that our results solves the problem of
computing the   large deviations function for negative $\Delta f$. One could therefore start to study more
difficult problems, like the large deviation function for positive  $\Delta f$ in the SK model. One could also
try to
extend our results to other models, such as Bethe lattices or large dimensions short range models, work is in progress in these directions \cite{PRprep}.

{\bf Aknowledgements} - We thank the Authors of Ref. \cite{CMPP} for giving us their numerical data.

\appendix

\section{Power Series of $\Phi(n)$}
In this appendix we report  the power series of $\Phi(n)$ of the SK model in the low temperature phase up to the 18th order in $n$ and $\tau=1-T$. At all order in $\tau$ the smallest power of $n$ is $n^5$ and there is no $n^6$ term.
\begin{widetext}
\beqa
\Phi(n) & = & - \frac{1}{4} -\ln 2 -\frac{\tau}{4}+\tau \ln 2 - \frac{{\tau}^2}{4} - \frac{{\tau}^3}{12} + \frac{{\tau}^4}{24} - \frac{{\tau}^5}{120} + 
  \frac{3\,{\tau}^6}{20} - \frac{79\,{\tau}^7}{140} + \frac{1679\,{\tau}^8}{560} - \frac{13679\,{\tau}^9}{720} + 
  \frac{1728361\,{\tau}^{10}}{12600} +
  \nonumber
  \\
& & -\frac{19214684\,{\tau}^{11}}{17325} + \frac{2741593487\,{\tau}^{12}}{277200} - 
  \frac{3939806687\,{\tau}^{13}}{40950} + \frac{773933492429\,{\tau}^{14}}{764400} - \frac{86662083146207\,{\tau}^{15}}{7567560} +
\nonumber
\\
& & + 
  \frac{139738065304401461\,{\tau}^{16}}{1009008000} - \frac{45875375549246420713\,{\tau}^{17}}{25729704000} + 
  \frac{11276190176083149262457\,{\tau}^{18}}{463134672000} +  
  \nonumber
  \\
& n^5 & \left( - \frac{9}{5120}   - \frac{99\,\tau}{5120} - \frac{27\,{\tau}^2}{320} - \frac{279\,{\tau}^3}{1280} - 
  \frac{981\,{\tau}^4}{2560} - \frac{351\,{\tau}^5}{400} + \frac{2799\,{\tau}^6}{12800} - 
  \frac{344241\,{\tau}^7}{22400} + \frac{47010861\,{\tau}^8}{358400}+ \right.
  \nonumber
  \\
& &
\left. - \frac{36684189\,{\tau}^9}{25600} + 
  \frac{830566899\,{\tau}^{10}}{51200} - \frac{1928757352257\,{\tau}^{11}}{9856000} + 
  \frac{98506298782713\,{\tau}^{12}}{39424000} - \frac{8635947355938261\,{\tau}^{13}}{256256000}\right)+
\nonumber
\\
& n^7 & 
\left( \frac{81}{143360} - \frac{2673\,\tau}{143360} - \frac{7047\,{\tau}^2}{35840} - \frac{35559\,{\tau}^3}{35840} - 
  \frac{75573\,{\tau}^4}{35840} - \frac{1943757\,{\tau}^5}{179200} + \frac{7442847\,{\tau}^6}{179200}+\right.
  \nonumber
\\  
& & \left.
-  \frac{762113853\,{\tau}^7}{1254400} + \frac{17011569051\,{\tau}^8}{2508800} 
 - \frac{210723811119\,{\tau}^9}{2508800} 
  +\frac{13663823711841\,{\tau}^{10}}{12544000} - \frac{1027400967213903\,{\tau}^{11}}{68992000}\right)+
\nonumber
\\
& n^8 & \left(
\frac{243}{32768} + \frac{4131\,\tau}{32768} + \frac{15309\,{\tau}^2}{16384} + \frac{34263\,{\tau}^3}{8192} + 
  \frac{429381\,{\tau}^4}{32768} + \frac{2740311\,{\tau}^5}{81920} + \frac{11253573\,{\tau}^6}{163840} +\right.
\nonumber
\\
& & \left. + 
  \frac{107945217\,{\tau}^7}{573440}
 - \frac{669127959\,{\tau}^8}{4587520} + \frac{12126319893\,{\tau}^9}{2293760} - 
  \frac{183401224893\,{\tau}^{10}}{3276800}
\right)+
\nonumber
\\
& n^9 & \left( 
-\frac{60021}{5734400}   - \frac{1720683\,\tau}{5734400} - \frac{3703563\,{\tau}^2}{1433600} - 
  \frac{48430143\,{\tau}^3}{2867200} - \frac{213993819\,{\tau}^4}{5734400} - \frac{2813451327\,{\tau}^5}{7168000} +\right.
\nonumber
\\
& & \left.
+ 
  \frac{2765750427\,{\tau}^6}{1146880} - \frac{1836578874951\,{\tau}^7}{50176000} + \frac{379740674928681\,{\tau}^8}{802816000} - 
  \frac{189083279254923\,{\tau}^9}{28672000}
\right)+
\nonumber
\\
& n^{10} & \left( 
\frac{155277}{3276800} + \frac{911979\,\tau }{819200} + \frac{36721917\,{\tau }^2}{3276800} + \frac{110699379\,{\tau }^3}{1638400} + 
  \frac{1837467099\,{\tau }^4}{6553600} + \frac{2973858543\,{\tau }^5}{3276800} 
+ \right.
  \nonumber
  \\
  & & \left.
+\frac{76627955097\,{\tau }^6}{32768000} + 
  \frac{104357662929\,{\tau }^7}{16384000} + \frac{853398339489\,{\tau }^8}{131072000}
\right)+
\nonumber
\\
& n^{11} & \left( 
- \frac{829433601}{5046272000}   - \frac{22603330989\,\tau }{5046272000} - \frac{7219643481\,{\tau }^2}{180224000} - 
  \frac{20133254457\,{\tau }^3}{57344000} - \frac{1149209550873\,{\tau }^4}{2523136000}
+ \right.
  \nonumber
  \\
  & & \left.
 - \frac{218994700592277\,{\tau }^5}{12615680000} + 
  \frac{1073108844538299\,{\tau }^6}{6307840000} - \frac{61281594304289307\,{\tau }^7}{22077440000}
\right)+
\nonumber
\\
& n^{12} & \left( 
\frac{131410269}{183500800} + \frac{2903445891\,\tau }{183500800} + \frac{4533651\,{\tau }^2}{25600} + \frac{22510325169\,{\tau }^3}{18350080} 
+ \right.
  \nonumber
  \\
  & & \left.
+ 
  \frac{1165811276367\,{\tau }^4}{183500800} + \frac{5189828163921\,{\tau }^5}{229376000} + \frac{89579196304317\,{\tau }^6}{917504000}
\right)+
\nonumber
\\
& n^{13} & \left( 
- \frac{15299148393873}{5651824640000}  - \frac{14089860473859\,\tau }{209924915200} - \frac{330886579531671\,{\tau }^2}{565182464000} - 
  \frac{13990469422488399\,{\tau }^3}{1836843008000} 
+ \right.
  \nonumber
  \\
  & & \left.
+ \frac{20015370592779843\,{\tau }^4}{1335885824000} - \frac{9384066047520578313\,{\tau }^5}{10496245760000}
\right)+
\nonumber
\\
& n^{14} & \left( 
\frac{66042560169}{5138022400} + \frac{580058908857\,\tau }{2569011200} + \frac{15101741931291\,{\tau }^2}{5138022400}
+ \right.
  \nonumber
  \\
  & & \left.
 + 
  \frac{125170590832281\,{\tau }^3}{6422528000} + \frac{7885818877083003\,{\tau }^4}{51380224000}
\right)+
\nonumber
\\
& n^{15} & \left( 
-\frac{1557529661529486369}{29389488128000000}   - \frac{13663672258178594727\,\tau }{14694744064000000} - 
  \frac{107529809054090820291\,{\tau }^2}{14694744064000000} 
+ \right.
  \nonumber
  \\
  & & \left.
- \frac{46100805957050412573\,{\tau }^3}{262406144000000}
\right)+
\nonumber
\\
& n^{16} & \left( 
\frac{190687314873528513}{723433553920000} + \frac{320621966627776497\,\tau }{180858388480000} + \frac{18912071856450181023\,{\tau }^2}{361716776960000}
\right)+
\nonumber
\\
& n^{17} & \left( 
- \frac{126373462658844234883011}{111915170791424000000}   - \frac{92659942781039607442731\,\tau }{55957585395712000000}
\right)+
\nonumber
\\
& n^{18} &
\frac{10254234479592769713}{1808583884800000}
\eeqa
\end{widetext}

\end{document}